# MoS$_2$/Al$_{0.68}$Sc$_{0.32}$N negative capacitance field-effect transistors

Seunguk Song,[1,†] Kwan-Ho Kim,[1,†] Srikrishna Chakravarthi,[1,†] Zirun Han,[1,2] Gwangwoo Kim,[1] Kyung Yeol Ma,[3] Hyeon Suk Shin,[3] Roy H. Olsson III,[1] and Deep Jariwala[1]*

[1]Department of Electrical and Systems Engineering, University of Pennsylvania, Philadelphia, Pennsylvania 19104, United States

[2]Department of Physics and Astronomy, University of Pennsylvania, Philadelphia, Pennsylvania 19104, United States

[3]Department of Chemistry, Ulsan National Institute of Science and Technology (UNIST), UNIST-gil 50, Ulsan 44919, Republic of Korea

[†]These authors equally contributed to this work: S. Song, K.-H. Kim, and S. Chakravarthi.

[*]Author to whom correspondence should be addressed: dmj@seas.upenn.edu

**ABSTRACT**

Al$_{0.68}$Sc$_{0.32}$N (AlScN) has gained attention for its outstanding ferroelectric properties, including a high coercive field and high remnant polarization. Although AlScN-based ferroelectric field-effect transistors (FETs) for memory applications have been demonstrated, a device for logic applications with minimal hysteresis has not been reported. This study reports on the transport characteristics of a MoS$_2$ negative capacitance FET (NCFET) based on an AlScN ferroelectric material. We experimentally demonstrate the effect of a dielectric layer in the gate stack on the memory window and subthreshold swing (SS) of the NCFET. We show that the hysteresis behavior of transfer characteristics in the NCFET can be minimized with the inclusion of a non-ferroelectric dielectric layer, which fulfills the capacitance-matching condition. Remarkably, we also observe the NC effect in MoS$_2$/AlScN NCFETs arrays based on large-area monolayer MoS$_2$ synthesized by chemical vapor deposition, showing the SS values smaller than its thermionic limit (~36-60 mV/dec) and minimal variation in threshold voltages (< 20 mV).

**MAIN TEXT**

The advancement of integrated circuit designs has heavily relied on the principles of Dennard scaling to maintain a constant power density as transistor dimensions continue to decrease.[1] However, as transistors reach nanoscale dimensions, reducing the supply voltage ($V_{dd}$) becomes difficult without increasing leakage current ($I_{off}$) and sacrificing on-state current ($I_{on}$).[1,2] Given that power density is directly proportional to $V_{dd}^2$, alternative approaches to further scale $V_{dd}$ are imperative to overcome this challenge of power dissipation.[3] A promising solution lies in overcoming the room-temperature (RT) thermionic limit of ~60 mV/dec for subthreshold swing (SS), as achieving a steeper switching behavior allows the transistor to maintain the same $I_{on}$ at lower $V_{dd}$ without an increase in $I_{off}$.[2] To overcome the thermionic limit, negative capacitance field-effect transistors (NCFETs) present a novel approach by incorporating a ferroelectric material in the gate stack. According to the Landau-Ginzburg-Devonshire (LGD) theory, ferroelectrics possess a metastable region in their polarization-dependent energy landscape that exhibits NC.[4,5] By introducing a ferroelectric insulator layer in the gate stack, SS can be reduced to less than 60 mV/dec at RT, as we can have $\frac{C_s}{C_{ins}} < 0$ in the following expression for SS:

$$SS = \frac{2.3 k_B T}{q} \left( \frac{\partial V_g}{\partial \psi_s} \right) = \frac{2.3 k_B T}{q} \left( 1 + \frac{C_s}{C_{ins}} \right) \qquad (1)$$

, where $\frac{2.3 k_B T}{q}$ is the thermionic limit at 300 K (~60 mV/dec), and $C_s$ and $C_{ins}$ are the semiconductor and gate insulator capacitance respectively.[6]

However, despite extensive research on ferroelectric materials and theoretical frameworks of the NC effect, achieving reliable hysteresis-free steep switching in NCFETs has been challenging.[4,5,7] The main difficulty is stabilizing the NC effect in the NCFET, which requires incorporating a conventional, non-ferroelectric dielectric as an interlayer in series with the ferroelectric material in the gate stack.[7] This creates a stable energy minimum in the NC region while reducing device hysteresis.[7] To eliminate hysteresis, the overall capacitance of the device must be positive, which can be achieved by ensuring that the magnitude of the ferroelectric capacitance ($|C_{FE}|$) remains higher than the combination of all remaining capacitances ($C_{REM}$).[7]

Using the LGD theory, one can obtain the following approximation for $|C_{FE}|$ of a ferroelectric material with a certain thickness ($t_{FE}$) as follows:[8, 9]

$$|C_{FE}| \approx \frac{2}{3\sqrt{3}} \frac{P_r}{E_c \cdot t_{FE}} \qquad (2)$$

As suggested by Equation (2), it is preferable to use a ferroelectric material with a large remnant polarization ($P_r$) and coercive field ($E_c$) to obtain a high $|C_{FE}|$.[8, 9] Furthermore, the S-shaped polarization-electric field (*P-E*) characteristics indicate that larger $E_c$ values could expand the NC region, which maintains the NC effect over a wide range of gate voltage ($V_g$) sweeps.[8, 10]

Recently discovered wurtzite-structured $Al_{1-x}Sc_xN$ (AlScN) possesses unique ferroelectric properties,[11-13] surpassing other thin-film ferroelectrics such as $Hf_xZr_yO_2$ and $Pb[Zr_xTi_{1-x}]O_3$ in $P_r$ and $E_c$.[14] AlScN exhibits substantially higher $P_r$ ranging from ~70 to 110 µC/cm² and $E_c$ between ~2 and 9 MV/cm, both of which can be adjusted by modifying the Sc doping levels.[11, 12] Consequently, AlScN provides a highly tunable platform for ferroelectric NC devices. Additionally, AlScN demonstrates excellent thermal stability of ferroelectric properties, maintaining its ferroelectricity over 1,000 °C.[14] It can also be deposited at low temperatures (< 300-400 °C) compatible with industry-standard Back-End-of-Line processes.[13] However, despite these desirable characteristics, no AlScN-based NCFETs have been reported yet.

Here, we report NCFETs using few-layer and monolayer $MoS_2$ channels and an $Al_{0.68}Sc_{0.32}N$ ferroelectric material in the gate-dielectric stack. $MoS_2$ is used for the semiconducting channel due to its near-constant channel capacitance enabling steep switching over large current ranges.[8] We studied the effect of the interlayer on the hysteresis and SS characteristics of AlScN-based NCFETs using various interlayer materials ($AlO_x$, $HfO_x$, and hexagonal BN (h-BN)) and thicknesses (0.5–2 nm). Notably, we observe $HfO_x$/AlScN-based NCFET devices with sub-thermionic SS value (~30.7 mV/dec) with a reduced shift in threshold voltage ($V_{th}$).

Figure 1a illustrates the cross-sectional schematic of our fabricated NCFETs (see also Figure S1). The device consists of a semiconducting channel made from mechanically exfoliated few-layer $MoS_2$, while a 45-nm-thick $Al_{0.68}Sc_{0.32}N$ layer deposited on Pt(111) functions as the ferroelectric gate insulator. In/Au is selected as S source and drain (S/D) contact electrodes to minimize the possibility of developing a degraded contact interface due to chemical interaction between MoS$_2$ and the contact.[15, 16] For the interlayer, HfO$_x$ and AlO$_x$ with various thicknesses are deposited, and h-BN is inserted via a wet transfer method (see Supplementary Methods for more details on device fabrication). Figure 1b shows a representative optical micrograph of the fabricated NCFETs with a few-layer MoS$_2$ channel (green) and In/Au contacts (gold).

Figure 1c shows transfer characteristics of two different FETs under the $V_g$ range from -12 V to 12 V; one with only the AlScN layer (blue) and the other with only a normal dielectric oxide layer (SiO$_2$/Si; orange). The AlScN-only FET is referred to as ferroelectric FET (FeFET), commonly used in memory applications due to its ability to provide a large memory window (MW) in the transfer characteristics, which corresponds to a shift in $V_{th}$ (e.g., MW of ~6.0 V in Figure 1c). However, the large MW is undesirable in digital logic devices, and thus we focus on minimizing MW (i.e., $V_{th}$ shift) later in this paper. The FeFET exhibits a counterclockwise hysteresis curve that is indicative of ferroelectric gating, as a non-volatile, reduced channel conductance remains at a high $V_g$ regime.[17] On the other hand, the FET with SiO$_2$ layer with a thickness of 50 nm (without AlScN) shows a clockwise hysteresis curve, mainly caused by charge traps at the dielectric/MoS$_2$ interface.[17]

In Figure 1d, the transfer characteristics of a FET incorporating both AlScN and a 1-nm-thick HfO$_x$ interlayer are presented. The HfO$_x$/AlScN FET demonstrates significantly reduced hysteresis in the switching behavior. The modulation of $V_{th}$ due to sweep direction is only ~0.7 V, which is smaller than that in a normal FeFET without any interlayer (~1.1 V in Figure S2). The smaller $V_{th}$ shift of HfO$_x$/AlScN FET can be attributed to the overall device capacitance being more positive and approaching the NCFET capacitance matching condition. The counterclockwise hysteresis curve observed in the HfO$_x$/AlScN device further confirms that the dominating switching behavior is attributed to the ferroelectricity of AlScN rather than the effect of interfacial charge traps.

The MoS$_2$/HfO$_x$/AlScN NCFET exhibits negative drain-induced barrier lowering (DIBL), since the $I_{ds}$ values are higher for $V_g$ values smaller than $V_{th}$ when measured at drain-source voltage ($V_{ds}$) of 0.1 V *vs.* 1.0 V (Figure 1d). This observation further supports our claim of observation of NC in these devices.[7, 18] In addition, the minimum SS (SS$_{min}$) achieved during the backward $V_g$ sweep (~63.9 mV/dec) is close to the thermionic limit of ~60 mV/dec at RT, as illustrated in the SS-$I_{ds}$ plot in Fig. 1e. The significantly lower $SS_{min}$ value compared to the AlScN-only FeFET (~209 mV/dec) and SiO$_2$-only FET (~1,558 mV/dec) indicates the stacked gate structure with matched capacitance effectively reduces the SS. Note that, the $SS_{min}$ can be even smaller (~46 mV/dec) when decreasing $V_{ds}$ step size from 50 mV to 5 mV (Figure S3), which proves the robust NC effect irrespective of sweep speeds.[19] Figure 1f presents the $I_{ds}$-$V_{ds}$ characteristics of the same MoS$_2$ NCFET. At high $V_g$ bias (5 V), negative differential resistance (NDR) is observed (Fig. 1g). The NDR effect is believed to be caused by the negative DIBL effect and/or the self-heating effect resulting from large drain current and voltage in the NCFET[7, 9, 18, 20, 21] providing additional support to our claims of an NCFET effect in these devices.

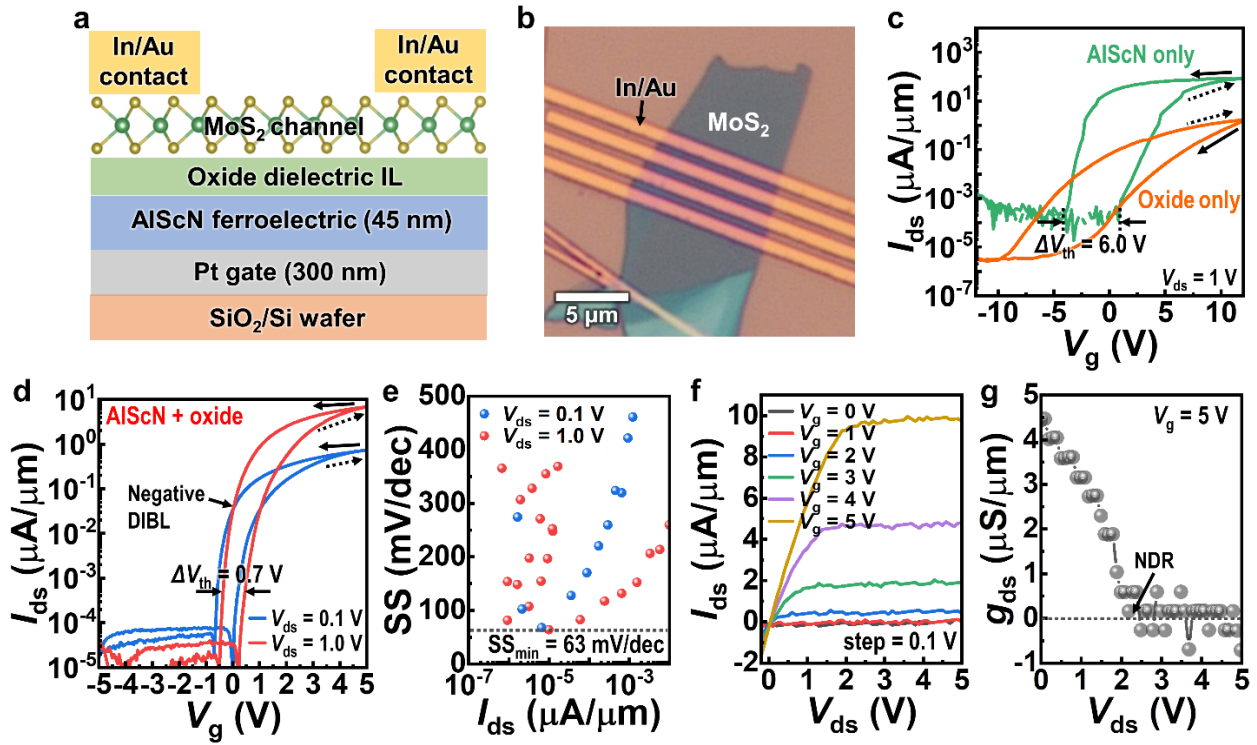

**FIG. 1.** (**a**) Schematic of MoS$_2$/AlScN NCFET with a dielectric interlayer. (**b**) Optical microscope image showing mechanically exfoliated MoS$_2$ along with patterned electrodes. (**c**) Semilogarithmic scale transfer characteristics of devices with only oxide dielectric (i.e., SiO$_2$/Si) and only ferroelectric AlScN layer (i.e., AlScN/Pt). Both curves measured at $V_{ds}$ = 1 V and $V_g$ from -12 to 12 V. (**d-g**) Electrical characterizations of capacitance matched NCFETs with AlScN and 1-nm-thick HfO$_x$ layer. (**d**) Semilogarithmic scale transfer characteristic of the NCFET, measured at $V_{ds}$ of 0.1 V (blue) and 1 V (red) ($V_{ds}$ step: 50 mV). (**e**) Corresponding SS vs. $I_{ds}$ plot under the backward $V_g$ sweep, indicating the $SS_{min}$ of ~63.9 mV/dec. (**f**) Output characteristic of MoS$_2$ NCFET with a $V_{ds}$ step size of 0.1 V. (**g**) Corresponding conductance ($g_{ds}$) depending on $V_{ds}$ at the $V_g$ of 5 V.

To comprehensively investigate the impact of dielectric interlayers on the performance of NCFETs, we have fabricated devices with mechanically exfoliated few-layer MoS$_2$ channels incorporating various interlayer materials and thicknesses. The corresponding transfer curves are presented in Figure 2a-f, allowing for a systematic understanding of how the thickness and type of the interlayer dielectric affect the transport characteristics of NCFETs. All the transfer curves clearly demonstrate the dependence of SS and MW on the type and thickness of the interlayer dielectric. Interestingly, in Figures 2a-f, a reduction in MW to values below ~500 mV is observed, regardless of the interlayer type. Notably, all curves exhibit the counterclockwise hysteric behavior, with an $I_{on}/I_{off}$ greater than ~10$^5$ indicating that the ferroelectric switching is actively occurring during $V_g$ sweep of the device.

We find that interlayers with higher dielectric constants result in steeper SS values at a constant thickness. Due to the high dielectric constant (i.e., $k$ of ~16-25) of HfO$_x$, it exhibits the lowest SS values, as indicated in Figures 2a-c. Interlayers composed of high-$k$ dielectrics are expected to exhibit a higher capacitance ($C_{IL}$). It is clear from Equation (1) that to achieve a sub-thermionic SS value, the factor $\left(1 + \frac{C_S}{C_{ins}}\right) = \left(1 + \frac{C_S}{C_{IL}} - \frac{C_S}{|C_{FE}|}\right)$ must be less than 1. Consequently, the SS of NCFET with a $|C_{FE}|$ is expressed as follows:

$$SS = \frac{2.3 k_B T}{q}\left(1 + \frac{C_S}{C_{IL}} - \frac{C_S}{|C_{FE}|}\right) \qquad (3)$$

Given Equations (2) and (3), to have a SS smaller than 60 meV/dec, the condition of $|C_{FE}| < C_{IL}$ must be met, and increasing the $C_{IL}$ assists in maintaining this relationship.[22] To this end, we successfully demonstrate a sub-thermionic SS$_{min}$ value of ~30.6 mV/dec for the 0.5 nm thick HfO$_x$ device, as shown in Figure 2a (the corresponding SS *vs.* $I_{ds}$ curve is shown in Figure S4.).

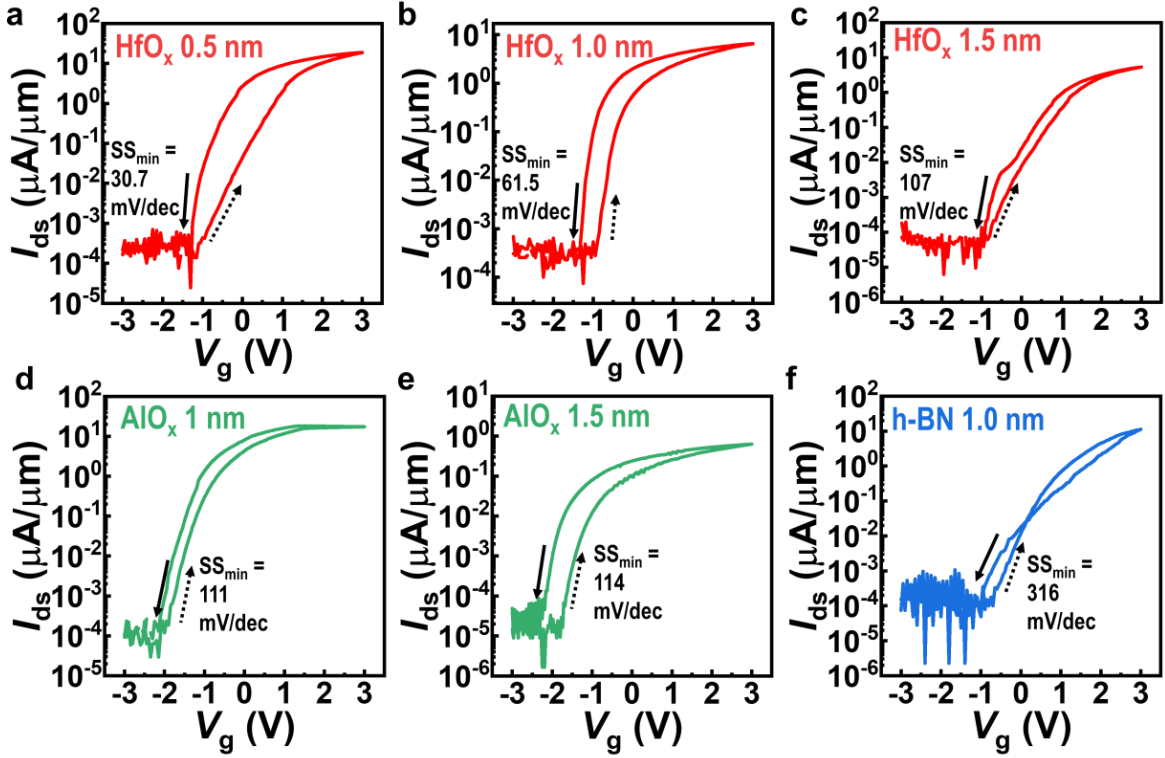

**FIG. 2.** Transfer characteristics for mechanically exfoliated MoS$_2$-based NCFETs with an interlayer of (**a**) 0.5 nm thick HfO$_x$, (**b**) 1 nm thick HfO$_x$, (**c**) 1.5 nm thick HfO$_x$, (**d**) 1.0 nm thick AlO$_x$, (**e**) 1.5 nm thick AlO$_x$, (**f**) 1 nm thick hexagonal BN (h-BN). All curves were measured at $V_{ds}$ = 1 V with a step size of 50 mV. The corresponding SS *vs.* $I_{ds}$ plots are shown in Figure S4.

Equation (2) also indicates that interlayers with greater thickness can lead to higher SS values due to the inverse relationship between $C_{IL}$ and dielectric thickness. Figure 3a illustrates how the SS extracted from our devices appropriately conforms to the expected trend based on the dielectric constant and thickness of the interlayer. Furthermore, Figure 3b demonstrates an excellent linear fit to a SS *vs.* $\frac{1}{C_{IL}}$ plot for the devices with the HfO$_x$ and h-BN interlayers. By extrapolating to the intercept of $\frac{1}{C_{IL}} = 0$, where $SS = \frac{2.3 k_B T}{q}\left(1 - \frac{C_S}{|C_{FE}|}\right)$, we can extract a range for $|C_{FE}|$ values by using appropriate values of $C_s$. To this end, we approximate the MoS$_2$ channel as a conventional dielectric as its charge screening length (~7 nm) is larger than the channel thickness of our device.[23] Accounting for the variability in the number of MoS$_2$ layers due to exfoliation and utilizing previously reported few-layer dielectric constants for the material,[24] we use $C_s$ values range of 2.53 to 3.79 µF/cm². This in turn allows us to extract a range of $C_{FE}$ values between -2.18 and -3.26 µF/cm². Note that, the extracted $C_{FE}$ of our 45-nm-thick AlScN (-2.18 to -3.26 µF/cm²) is lower in magnitude compared to the calculated $C_{FE}$ of the 20-nm-thick ferroelectric $Hf_xZr_yO_2$ (~ -13.1 µF/cm²);[7] however, the NC value for AlScN/MoS$_2$ system will likely be enhanced as the AlScN layer is scaled to lower thickness ($t_{FE}$) with an accompanying increase in $E_c$ in accordance with Equation (2).

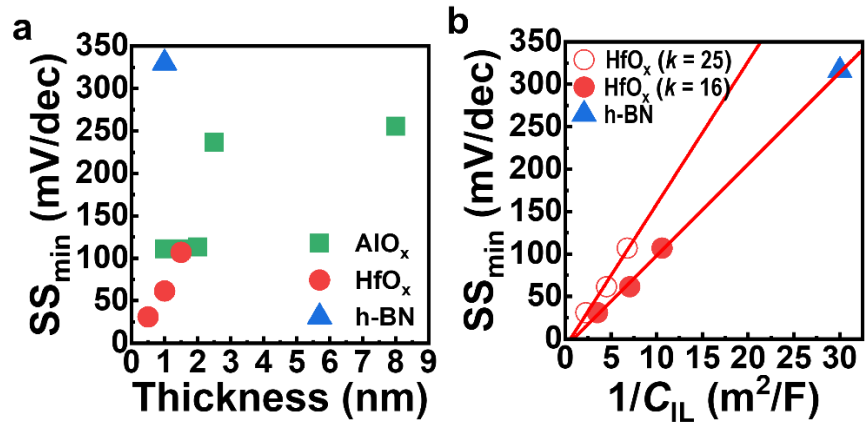

**Fig. 3.** (**a**) Extracted SS$_{min}$ values as a function of interlayer type and thickness. (**b**) SS$_{min}$ vs. $\frac{1}{C_{IL}}$ scatter plot. To account for variability in the dielectric constant of HfO$_x$, $C_{IL}$ values are calculated assuming $k = 16$ and $k = 25$.

As mechanical exfoliation is unsuitable for large-area applications, we additionally fabricated FeFETs and NCFETs utilizing large-area monolayer MoS$_2$ grown through chemical vapor deposition (CVD), as depicted in Figure 4a (and Figure S5a). The device configuration is the same as those using mechanically exfoliated flakes to ensure its comparable capacitance network (see Figure S1 and Supplementary Methods). The transfer curves of the MoS$_2$/AlScN FeFET devices (without dielectric interlayers) exhibit sub-thermionic SS switching behavior (Figure 4b). The minimum observed SS$_{min}$ values of the six devices are calculated (Figure 4c) and listed (Figure 4d), and the steepest SS is confirmed to be ~36 mV/dec. Importantly, all six devices demonstrate steep $SS_{min}$ values, which are the same or lower than the thermionic limit (< 60 mV/dec).

On the other hand, CVD-MoS$_2$ NCFET with a 2 nm thick interlayer of AlO$_x$ results in a nearly hysteresis-free transfer curve (Figure 4e) with an $SS_{min}$ of ~90 mV/dec (Figure 4f). Additionally, the NCFET exhibits an $I_{on}/I_{off}$ exceeding ~10$^5$, despite the relatively narrow $V_g$ sweep range of -3V to 3V, even with a total gate stack thickness of ~47 nm (~2 nm AlO$_x$/~45 nm AlScN). Similar to the observations in the mechanically exfoliated samples, thicker dielectric interlayers lead to reduced hysteresis, though maintaining a steep SS below 60 mV/dec becomes challenging. The AlO$_x$/AlScN-based NCFET follows a similar trend, necessitating a 2 nm thick dielectric layer to achieve minimal hysteresis (~20 mV).

As an interlayer, HfO$_x$ can surpass AlO$_x$ in achieving steeper SS in NCFETs at the same thickness due to higher $k$. By utilizing the calculated $C_{FE}$ from Figure 3, we determine the optimal thickness of the HfO$_x$ interlayer to be 0.9 nm using the capacitance-matching condition, ensuring minimal hysteresis in the CVD-MoS$_2$ NCFET. The representative transfer curve and SS-$I_{ds}$ plot of NCFET with a 0.9 nm-thick HfO$_x$ are shown in Figure 4g and Figure 4h, respectively. This specific NCFET exhibits a small MW of ~0.34 V and an $SS_{min}$ of ~51 mV/dec. Notably, the extracted $SS_{min}$ (~51 mV/dec) is smaller than that with the AlO$_x$ interlayer (~90 mV/dec in Figure 4e) and the RT-thermionic limit (~60 mV/dec). Moreover, this NCFET exhibits negative DIBL and NDR effects, confirming the occurrence of the NC effect (Figure S5). However, not all NCFETs display sub-thermionic $SS_{min}$; for example, the averaged MW and $SS_{min}$ of the NCFETs with a 300-nm-long channel are ~69.7 ± 12.1 mV/dec and ~0.26 ± 0.15 V (average ± standard deviation), possibly due to non-uniformity of channel quality and MoS$_2$/AlScN interface (discussed later).

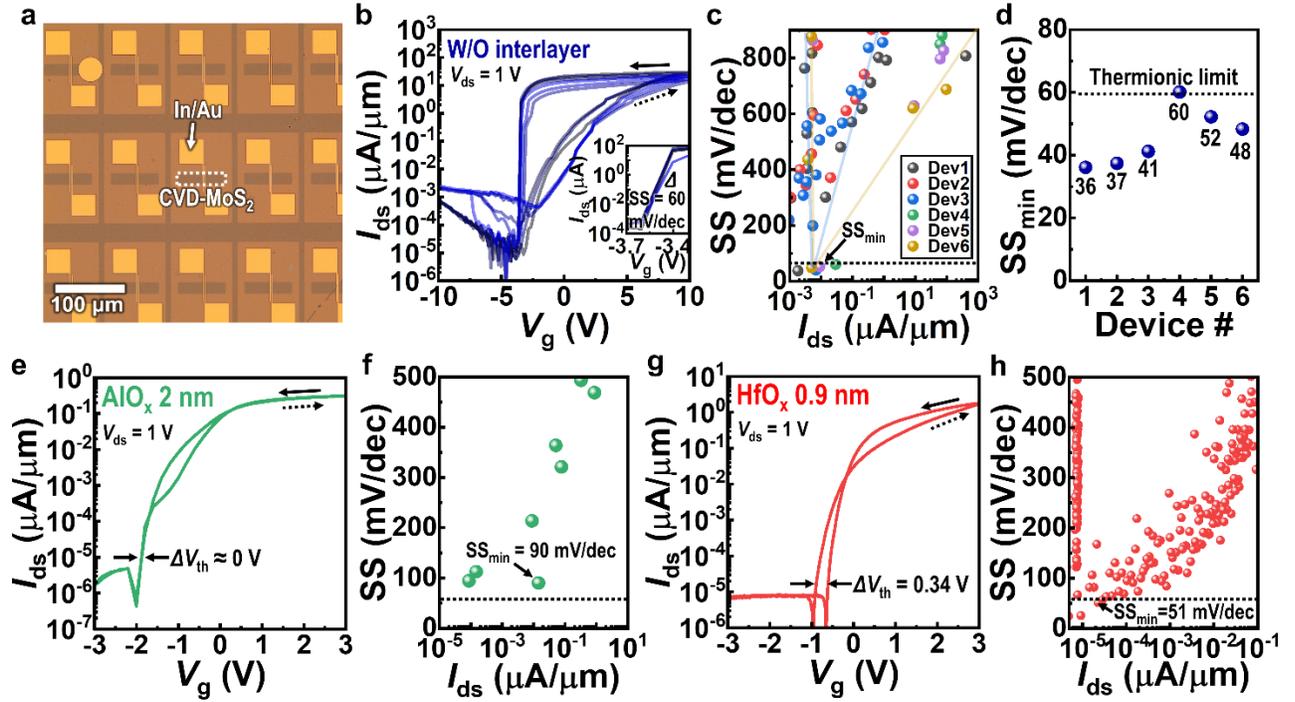

**FIG. 4.** (**a**) Optical microscope image of CVD MoS$_2$ NCFET array. (**b**) Transfer characteristics for various devices without interlayer ($V_{ds}$ = 1 V). The inset shows the zoomed-in curves to highlight their SS values smaller than 60 mV/dec. (**c**) Corresponding SS-$I_{ds}$ plot and (**d**) summary of SS of the devices without interlayer, displaying sub-60 mV/dec SS values. (**e**) Transfer characteristics for a device with 2 nm AlOx interlayer, and its (**f**) SS values depending on the $I_{ds}$. (**g**) $I_{ds}$-$V_g$ plot of representative NCFET with 0.9 nm thick HfO$_x$ interlayer, and (**h**) its corresponding SS *vs.* $I_{ds}$, demonstrating the SS$_{min}$ of ~51 mV/dec.

To analyze device-to-device variations and channel length ($L$)-dependence on SS$_{min}$, the transfer characterization of NCFETs with varying $L$ (300-1,500 nm) is conducted. Figures 5a and b display the overlayed transfer curves of 20 devices during the forward and backward sweep of $V_g$ respectively. As expected, the $I_{on}$ is highest when the $L$ is shortest (~300 nm) (Figure 5c). The best $I_{on}$ approaches ~9.0 µA/µm at $V_{ds}$ = 1 V (and ~0.91 µA/µm at $V_{ds}$ = 0.1 V). Remarkably, this high $I_{on}$ is comparable to those of previously reported Hf$_x$Zr$_y$O$_2$-based NCFETs with the mechanical exfoliated MoS$_2$[7, 25-28] and WSe$_2$[29, 30] multilayers (~10$^{-2}$-2 µA/µm at $V_{ds}$ = 0.1 V) (Table S1). We also observe that the SS$_{min}$ values increase over $L$ for our MoS$_2$/HfO$_x$/AlScN NCFETs (Figure 5d). Interestingly, this is contrary to the behavior for MoS$_2$ MOSFETs (e.g., increase or unchanged SS$_{min}$ as $L$ decreases),[31, 32] indicating that the transport characteristics are largely influenced by the NC effect rather than dielectric capacitances. However, sub-thermionic SS$_{min}$ is only observed in three devices with a short $L$ of ~300 or ~600 nm (red points below the dashed line in Figure 5d). The CVD-grown large-area MoS$_2$ possesses crystalline imperfections such as grain boundaries and sulfur vacancies, which may lead to defect-mediated variations on the channel (e.g., decreased $C_s$, increased interfacial trap density, and field-induced defect migration),[33-36] thus altering SS$_{min}$. Furthermore, the degradation of SS$_{min}$ can be substantial as $L$ increases due to an increased number of line defects in longer CVD-MoS$_2$ channels.[35] This suggests that using a shorter channel and maintaining a higher crystalline quality of 2D channel are key to achieving reproducible high performance in NCFETs.

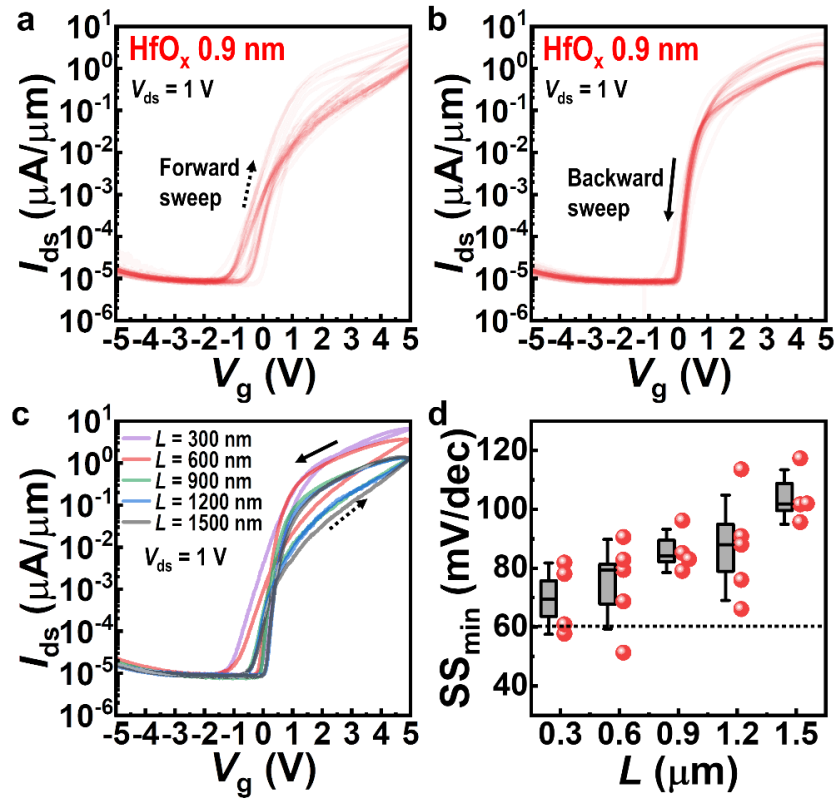

**FIG. 5.** (**a, b**) The overlayed transfer curves of 20 different CVD-MoS$_2$/HfO$_x$/AlScN NCFETs with a varying $L$ (300-1,500 nm) recorded during (**a**) forward sweep and (**b**) backward sweep of $V_g$ ($V_{ds}$ = 1 V). (**c**) Representative transfer characteristics of the NCFETs with different $L$, showing the counterclockwise hysteresis behavior. (**d**) Boxplot of SS$_{min}$ depending on the $L$, where the statistics show the median (lines in the boxes), standard deviations (error bars), and standard errors (boxes).

In summary, we present the experimental realization of 2D NCFETs utilizing AlScN as the ferroelectric layer and MoS$_2$ as the channel. Sub-thermionic SS$_{min}$ values as low as ~30.7 mV/dec, as well as negative DIBL in the transfer characteristics and an NDR effect in output characteristics, were successfully observed, providing clear evidence of the NC effect. By modulating the interlayer dielectric material and its thickness, we are able to reduce hysteresis and MW while simultaneously maintaining a steep SS. Our findings are a first step towards the application of AlScN, a versatile and tunable ferroelectric material, in steep-switching logic devices. Future work on these devices will involve optimizing AlScN composition and thickness together with interlayers for lower operating voltage NCFETs with sub-thermionic SS values for a large swing in current magnitudes.

## SUPPLEMENTARY MATERIAL

See the supplementary material for Supplementary Methods and Figure S1-5.

## ACKNOWLEDGEMENT

D.J. and S.S. acknowledge support from AFOSR GHz-THz program grant number FA9550-23-1-0391. S.S. also acknowledges partial support for this work by Basic Science Research Program through the National Research Foundation of Korea (NRF) funded by the Ministry of Education (Grant No. 2021R1A6A3A14038492).

## AUTHOR DECLARATIONS

### Conflict of Interest

The authors declare that they have no conflicts of interest.

### Author Contributions

D. J. supervised the study. S.S. and K.-H.K. fabricated the devices. S.S., K.-H.K., and S.C. performed the measurements. Z.H. and G.K. contributed to the discussion. K.Y.M. and H.S.S.

grew and provided h-BN layers. R.H.O. contributed to the preparation of AlScN. S.S., K.-H. K. and S. C. wrote the manuscript.

## DATA AVAILABILITY

The data that support the findings of this study are available from the corresponding author upon reasonable request.

Supplementary material

# MoS$_2$/Al$_{0.68}$Sc$_{0.32}$N negative capacitance field-effect transistors

Seunguk Song,[1,†] Kwan-Ho Kim,[1,†] Srikrishna Chakravarthi,[1,†] Zirun Han,[1,2] Gwangwoo Kim,[1] Kyung Yeol Ma,[3] Hyeon Suk Shin,[3] Roy H. Olsson III,[1] and Deep Jariwala[1]*

[1]*Department of Electrical and Systems Engineering, University of Pennsylvania, Philadelphia, Pennsylvania 19104, United States*

[2]*Department of Physics and Astronomy, University of Pennsylvania, Philadelphia, Pennsylvania 19104, United States*

[3]*Department of Chemistry, Ulsan National Institute of Science and Technology (UNIST), UNIST-gil 50, Ulsan 44919, Republic of Korea*

[†]*These authors equally contributed to this work: S. Song, K.-H. Kim, and S. Chakravarthi.*

*Author to whom correspondence should be addressed: dmj@seas.upenn.edu

- Supplementary Methods
- Figures S1-5
- Supplementary References

## SUPPLEMENTARY METHODS

### Fabrication of interlayer/AlScN/Pt sample

A 45 nm thick layer of $Al_{0.68}Sc_{0.32}N$ was deposited on pre-deposited Pt/SiO$_2$/Si using a pulsed-DC reactive sputtering deposition system (Evatec, CLUSTERLINE 200 II pulsed DC PVD).[1, 2] The co-sputtering process involved separate 4-inch Al and Sc targets, operated at 1250 W and 698 W, respectively, at a chuck temperature of 350 °C with 10 sccm of Ar gas flow and 25 sccm of N$_2$ gas flow under the constant pressure of ~$1.45 \times 10^{-3}$ bar. The growth of AlScN was facilitated by the preferentially oriented (111) Pt on SiO$_2$/Si substrates, which promoted highly textured ferroelectricity along the [0001] direction. Following the preparation of AlScN/Pt/SiO$_2$/Si, oxide interlayers (ILs) were deposited using atomic layer deposition (ALD; Cambridge Nanotech S200 ALD). HfO$_x$ and AlO$_x$ ILs were grown using the precursors of $[(CH_3)_2N]_4Hf$ and $Al_2(CH_3)_6$, respectively, along with the H$_2$O precursor at a temperature of 150 °C. The thickness of the ILs was controlled through the ALD cycles, considering their deposition rate of ~0.9 Å/cycle for each oxide. The 3nm-thick-h-BN film was used as one of the interlayers in NCFETs, which was grown on a *c*-plane sapphire substrate by a low-pressure chemical vapor deposition (CVD) method using ammonia borane powders as the precursor. The previous report[3] contains the specific experimental details for growth and the subsequent polymethyl methacrylate (PMMA)-wet transfer process for h-BN.

### Fabrication of mechanically exfoliated MoS$_2$ NCFETs

The align markers were deposited on the IL/AlScN/Pt/SiO$_2$/Si sample using photolithography (SUSS MicroTec MA6 Gen3 Mask Aligner) and e-beam evaporation (Lesker PVD75 E-beam Evaporator) of Ti/Au with a thickness of 10/40 nm. Note that, the heat treatment up to 180 °C for 5 min was performed during the photoresist coating of LOR3A and S1813. Following that, mechanical exfoliation of MoS$_2$ flakes was carried out, and annealing under Ar gas at 180 °C for 1 h was conducted to ensure proper adhesion between MoS$_2$ and the substrate. Subsequently, the source and drain (S/D) contacts of In/Au (10/40 nm) were defined using e-beam lithography (EBL; Elionix ELS-7500EX) and thermal evaporation (Kurt J. Lesker Nano-36). The use of PMMA-based e-beam resist required a thermal budget process at 180 °C for 10 min. To prevent any high-energy process during the metal deposition process, the deposition rate was maintained below 0.03 Å/s. The Pt layer beneath the AlScN served as a gate contact, and the gate contact exposure was achieved by wet etching of AlScN through immersion in a KOH solution after fabricating the In/$MoS_2$/IL/AlScN/Pt heterostructure. No additional heat treatment was conducted after the device fabrication.

**Fabrication of CVD-grown $MoS_2$ NCFETs**

A monolayer $MoS_2$ thin film, grown via CVD on a (0001) *c*-cut sapphire substrate (purchased from 2D Semiconductors) was used for the NCFETs. The grain size of $MoS_2$ was random (~1-20 μm) and nearly continuous across the 1 x 1 $cm^2$. CVD-grown $MoS_2$ was transferred onto the oxide/AlScN/Pt/$SiO_2$/Si sample using a KOH-based wet process. A polymeric supporting layer of PMMA was coated onto the $MoS_2$/sapphire and then floated on a 0.1 M KOH solution to separate the $MoS_2$ from the sapphire substrate. The detached PMMA/$MoS_2$ free-standing layer was rinsed multiple times in distilled ionized water to prevent any damage to the AlScN caused by residues from the KOH-based process. Following the transfer of PMMA/$MoS_2$ onto the AlScN layer, the PMMA was removed using acetone. Subsequently, the channel-defining process was performed using EBL and a reactive ion etcher (RIE) with oxygen gas (Jupiter II RIE Plasma Etcher). The S/D electrodes were then deposited after the channel definition, utilizing the same method employed for mechanically exfoliated $MoS_2$ NCFETs. Etching the $MoS_2$ channel before depositing In/Au electrodes results in the same device capacitance network as when using a mechanically exfoliated flake. In this case, the contact pads are in direct contact with the dielectric interlayer surface, without any $MoS_2$ underneath (see Figure S1).

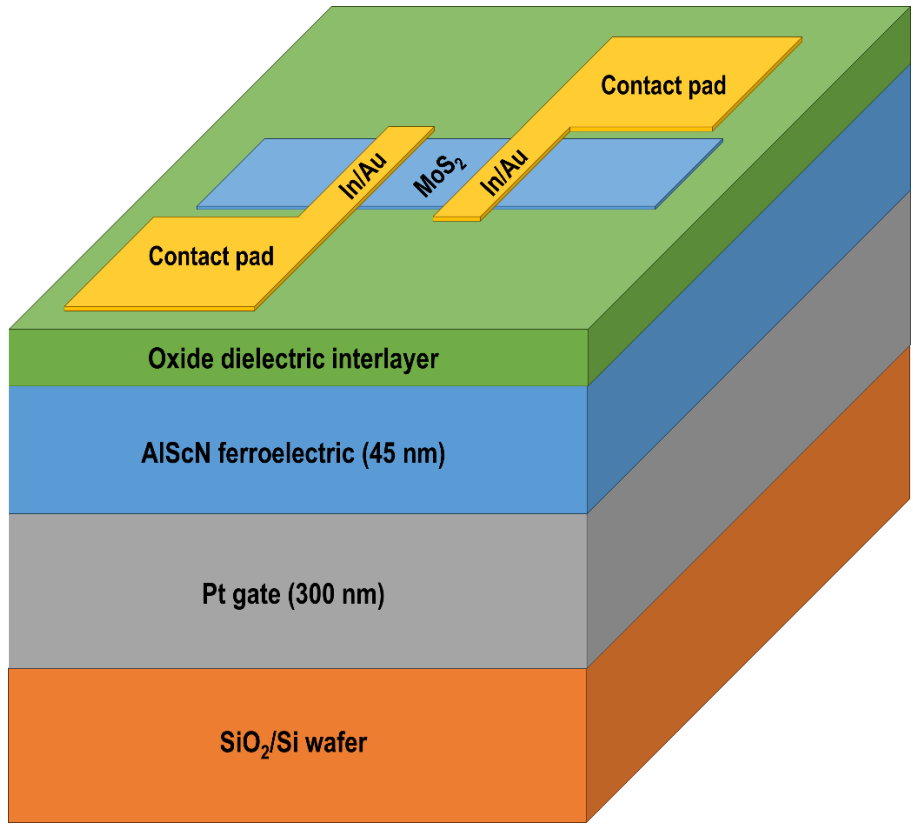

**Figure S1**. Schematic of MoS$_2$/AlScN NCFET with a dielectric interlayer. The contact pads are directly deposited onto the oxide interlayer surface. NCFETs with mechanically exfoliated flakes and CVD-grown MoS$_2$ have the same device configuration as depicted. Note that, the device configuration described differs from the one previously reported for the MoS$_2$/AlScN FeFET case.[4] In this particular instance, the MoS$_2$ layer exists below the contact pad, which is a result of the channel defining process after S/D deposition.

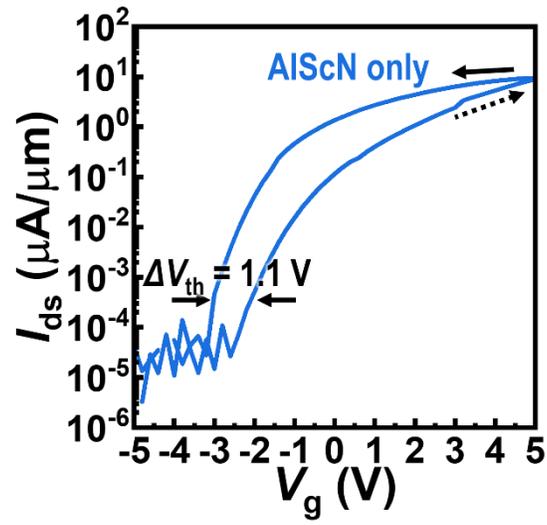

**Figure S2.** Semilogarithmic scale transfer characteristic of MoS$_2$/AlScN FeFET without any interlayers, measured at $V_{ds}$ of 1 V and $V_g$ range of -5 to 5 V.

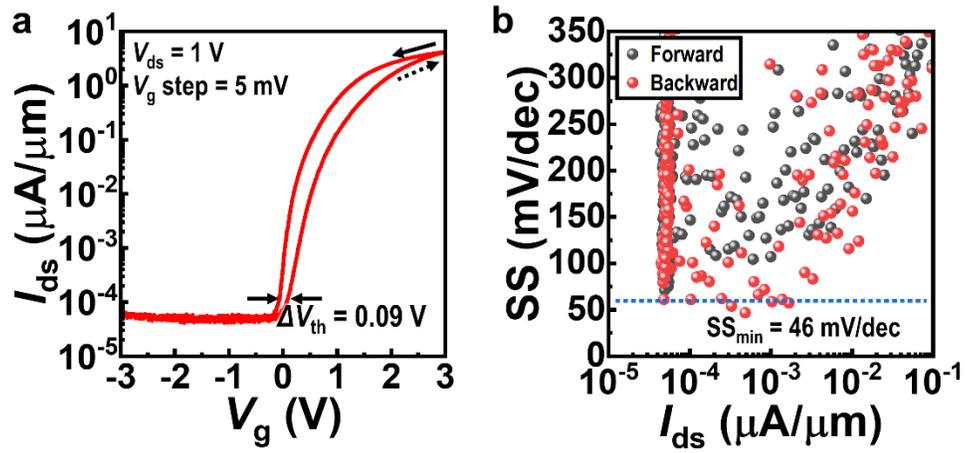

**Figure S3.** (**a**) Transfer characteristic of MoS$_2$ NCFET with a small $V_g$ step size of 5 mV. The 1.0 nm HfO$_x$ IL is used for capacitance matching. (**b**) Corresponding SS-$I_{ds}$ plot showing the SS$_{min}$ of ~46 mV/dec.

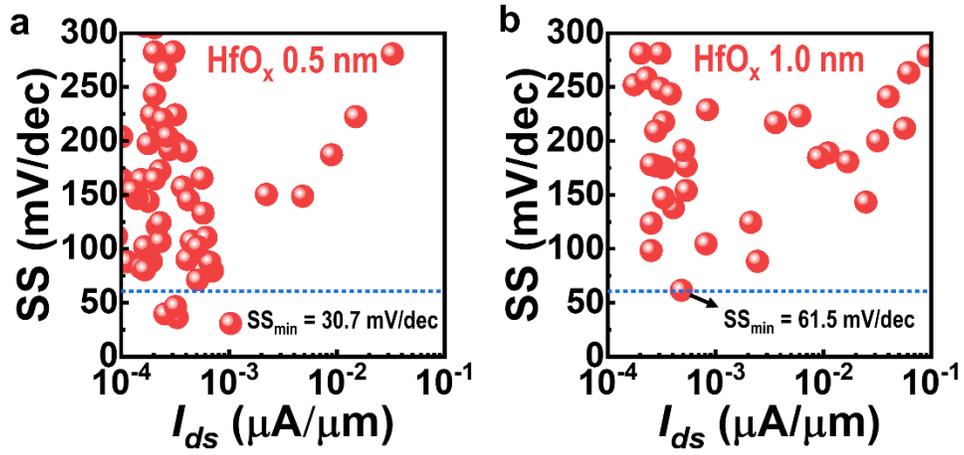

**Figure S4.** SS vs $I_{ds}$ curves of MoS$_2$ with (**a**) 0.5 and (**b**) 1.0 nm thickness HfO$_x$ interlayer. The corresponding transfer curves are presented in Figures 2a, b.

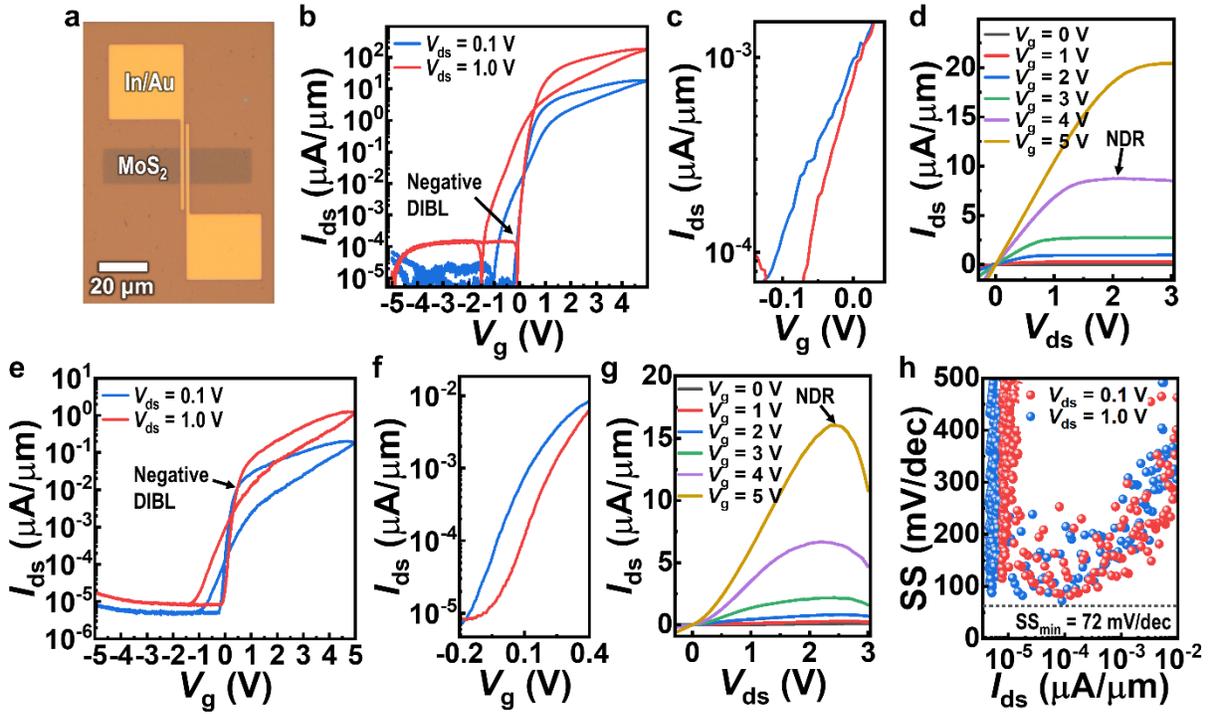

**Figure S5.** (**a**) OM image of a CVD-MoS$_2$ NCFET with a HfO$_x$ 0.9 nm interlayer. (**b, c**) Transfer curves of the NCFET for a device shown in Fig. 4g, h in the main text, and (**d**) its output characteristics. (**e-h**) Electrical transport of a MoS$_2$/HfO$_x$/AlScN NCFET different from that introduced in the main text, showing (**e**) transfer curve, (**f**) zoomed-in transfer curve to highlight the negative DIBL, (**g**) output curve and (**h**) SS-$I_{ds}$ plot.

**Table S1. Benchmarks of NCFETs with 2D semiconductor channels.** For NCFETs with 2D channel, the 3D ferroelectric gate of CuInP$_2$S$_6$ (CIPS), and Hf$_{1-x}$Zr$_x$O$_2$ (HZO) were used.

| Channel | Gate dielectric or ferroelectric | Interlayer | Contact metal | $SS_{min}$ (mV/dec) | $\Delta V_{th}$ (mV) | $L_{ch}$ (nm) | $I_{on}$ (μA·μm) at $V_{ds}$ = 0.1 V | $I_{on}/I_{off}$ ratio | Ref. |
|---|---|---|---|---|---|---|---|---|---|
| **CVD, monolayer MoS$_2$** | **AlScN** | **HfO$_x$ 0.9 nm** | **In/Au** | **51** | **340** | **300** | **0.91** | **10$^6$** | **This study** |
| Mechanically Exfoliated MoS$_2$ | CIPS | h-BN 7.5 nm | Cr/Au | 28 | 98 | 5700 | 1.96 (at $V_{ds}$ = 0.5V) | 10$^7$ | 5 |
| Mechanically Exfoliated MoS$_2$ | HZO | AlO$_x$ 2 nm | Ti/Au | 17.2 | 20 | 83 | 2 | 5 * 10$^6$ | 6 |
| Mechanically Exfoliated MoS$_2$ | HZO | AlO$_x$ 2 nm | Au | 23 | 77 | 2600 | 2 | 10$^8$ | 7 |
| Mechanically Exfoliated WSe$_2$ | HZO | AlO$_x$ 4 nm | Pt | 18.2 | 20 | 5000 | 1 | 10$^6$ | 8 |
| Mechanically Exfoliated MoS$_2$ | HZO | AlO$_x$ 2 nm | Ni | 5.6 | 12 | 2000 | 1 | 10$^7$ | 9 |
| Mechanically Exfoliated MoS$_2$ | HZO | AlO$_x$ 6nm | Cr/Au | 17.64 | Negligible | 3000 | 0.8 | 10$^7$ | 10 |
| Mechanically Exfoliated WSe$_2$ | HZO | AlO$_x$ 2 nm | Pt/Au | 40.2 | ~200 | 500 | 10$^{-2}$ | >10$^5$ | 11 |
| | HfO$_x$/Ni/AlO$_x$/HZO (10/20/3/20 nm) | | | 14.4 | ~120 | 1000 | 10$^{-2}$ | 10$^5$ | |

**SUPPLEMENTARY REFERENCES**